\documentclass[11pt]{article}

\usepackage[utf8]{inputenc}
\usepackage{amsmath, amssymb}
\usepackage{graphicx}
\usepackage{hyperref}
\usepackage{authblk} 
\usepackage{cite}
\usepackage[a4paper, left=1.5cm, right=1.5cm, top=2.5cm, bottom=2.5cm]{geometry} 

\title{Comparison of Different Rydberg Atom-Based Microwave Electrometry Techniques}

\author[1]{Eliel Leandro Alves Junior}
\author[1]{Manuel Alejandro Lefr{\'a}n Torres}
\author[2]{Jorge Douglas Massayuki Kondo}
\author[1]{Luis Gustavo Marcassa}

\affil[1]{ Instituto de F\'{i}sica de S\~{a}o Carlos, Universidade de S\~{a}o Paulo, Caixa Postal 369, 13560-970, S\~{a}o Carlos, SP, Brasil.}
\affil[2]{ Departamento de Física, Universidade Federal de Santa Catarina, Florianópolis 88040-900, SC, Brasil.}

\date{\today}

\begin{document}

\maketitle

\begin{center}
\textbf{Correspondence email}: \href{mailto:marcassa@ifsc.usp.br}{marcassa@ifsc.usp.br}
\end{center}

\begin{abstract}
In this study, we have compared different Rydberg atom-based microwave electrometry techniques under the same experimental conditions and using the same Rydberg states ($68S_{1/2}$, $68P_{3/2}$ and $67P_{3/2}$). The comparison was carried out for the following techniques: i) Auxiliary microwave field, ii)  Microwave amplitude modulation, and iii) Polarization spectroscopy. Our results indicate that all three techniques have a similar minimum measurable microwave electric field. A slightly better result can be obtained by performing polarization spectroscopy using a Laguerre-Gauss coupling laser beam.
\end{abstract}


\section{Introduction}
Rydberg atoms, characterized by high excitation levels and a significant principal quantum number $n$ \cite{gallagher2006rydberg, adamsry}, possess remarkable properties such as enhanced sensitivity to electromagnetic fields. Electromagnetically induced transparency (EIT) is a quantum interference effect \cite{fleischhauer2005electromagnetically, marangos1998electromagnetically} that creates a transparency window in a medium's absorption spectrum, facilitating precise and targeted detection. The synergy between Rydberg atoms and EIT has created an exciting research domain with various \cite{adams98, adams2019rydberg, Carr:12, otterbach}. In particular, in recent years, the application of microwave field detection (MW) has been successfully realized by numerous research teams \cite{sedlacek2012microwave, holloway2017electric, holloway2014broadband, holloway2014sub, PhysRevLett.111.063001, tanasittikosol2011microwave, fan2015atom, PhysRevApplied.5.034003, PhysRevA.90.043419, simons2016simultaneous, holloway2018quantum, simons2018electromagnetically, meyer2018digital, Fan:14}. Compact devices like atomic vapor cells have achieved unprecedented sensitivity and speed in detecting microwave signals.

The Rydberg atom-based microwave electrometry involves a four-level atom ladder configuration coupled by two laser beams (probe and coupling) and microwave radiation. Under particular conditions, the EIT transparency spectrum presents an Autler-Townes (AT) splitting \cite{cohen1996autler, anisimov2011objectively, abi2010electromagnetically}, which is very sensitive to the strength and frequency of the MW field. The AT splitting provides an accurate means to measure the MW electric field strength and is also traceable to the International System of Units (SI) \cite{sedlacek2012microwave}. Although Rydberg EIT sensing has led to notable advancements, it still faces certain challenges. The AT splitting defines the threshold for detecting minimal MW electric fields, constrained by the EIT linewidth \cite{holloway2017electric}. In observing the reduction of probe light transparency near the Rydberg EIT resonance, the smallest MW electric field strength detected to date in vapor cells is 55 nV/cm \cite{superheterodyne}. This approach offers high sensitivity but necessitates comparing transmittance measurements both with and without MW fields. Therefore, it is timely to develop techniques to measure small MW fields. 

Several techniques have been developed recently to lower the minimum detected MW field using simple spectroscopical methods. In a recent work, Jia and co-workers developed the auxiliary microwave field technique \cite{auxiliary}. It consists of an extra MW field in resonance with a nearby different Rydberg state, which is called the auxiliary field. This field introduces an AT splitting that is larger than the EIT linewidth. The measurement is performed in the target MW field, which is resonant to a different Rydberg state. The final AT splitting depends on both fields, and the authors were able to measure the field as low as 31 $\mu$V/cm at 14.2 GHz. In another work, Liu and co-workers have used MW amplitude modulation to reach a minimum detectable electric field strength of 430 $\mu$V/cm at 14 GHz \cite{MWmodulation}. Gomes et al. have investigated Rydberg-atom-based microwave electrometry using polarization spectroscopy and were able to detect 870 $\mu$V/cm at 11.6 GHz \cite{DuarteGomes_2024}. By combining a MW lens, the minimum field was improved to 310 $\mu$V/cm.

All these techniques were applied to different Rb atomic states, under different experimental conditions (probe and control laser powers). Therefore, its comparison is not straightforward. In this work, we have applied all these techniques (auxiliary microwave field \cite{auxiliary}; microwave amplitude modulation \cite{MWmodulation}; and polarization spectroscopy \cite{DuarteGomes_2024}) to the same Rydberg state under the same experimental conditions. Our results indicate that the best technique is polarization spectroscopy \cite{DuarteGomes_2024}. The structure of this work is organized as follows: i) Section \ref{Material} presents the experimental setup and procedure; ii) Section \ref{Results} provides the experimental results; iii) Section \ref{Discussion} presents the discussions; iv) Finally, Section \ref{Conclusion} concludes the study with the presentation of the main findings and conclusions.

\section{Materials and Methods}\label{Material}

\begin{figure}
      \centering
    \includegraphics[width=1\columnwidth]{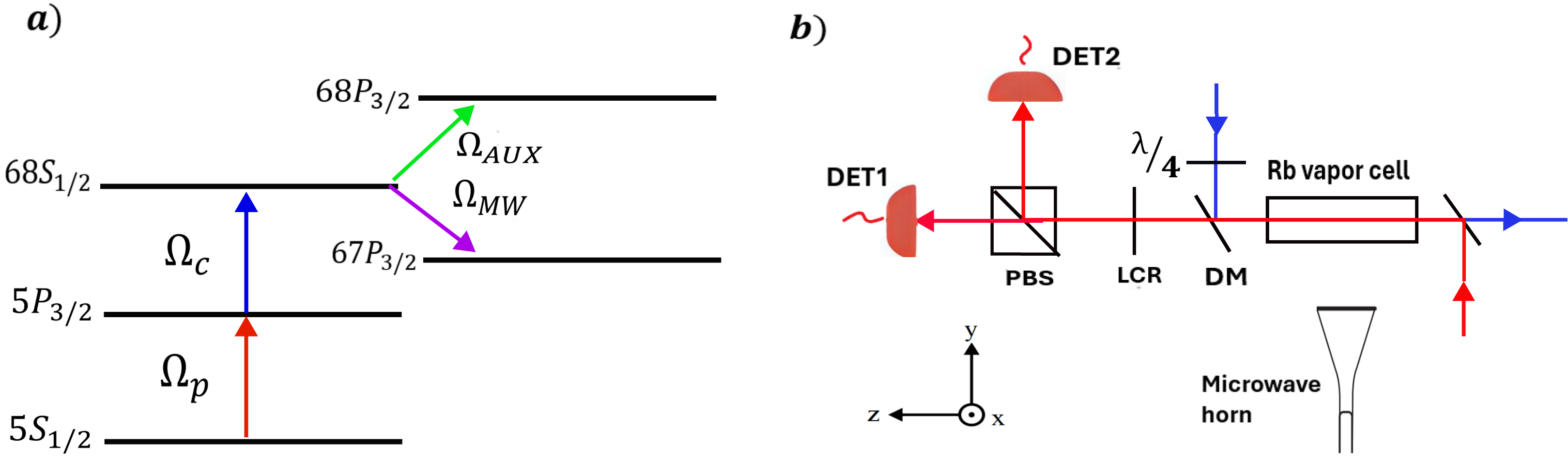}
    \caption{(Color online) (a) Scheme of the five-level atom. Levels $5S_{1/2}$ and $5P_{3/2}$ are coupled by a probe laser, levels $5P_{3/2}$ and $68S_{1/2}$ are coupled by a coupling laser. The levels $68S_{1/2}$ and $67P_{3/2}$ are Rydberg levels connected by MW radiation ($\Omega_{MW}$). The levels $68S_{1/2}$ and $68P_{3/2}$ are Rydberg levels connected by auxiliary MW radiation ($\Omega_{Aux}$). (b) Experimental setup:  waveplate ($\lambda/4$); polarizing beam splitter (PBS); Dichroic mirror (DM); liquid crystal retarder (LCR); vapor cell and photodetectors (DET1 and DET2).  A probe (red) and coupling (blue) laser beams counterpropagate inside a rubidium vapor cell under the influence of an MW field generated at the MW horn antenna. The probe laser beam is linearly polarized in the x-direction, while the MW electric field is polarized in the z-direction. For the EIT measurement, the coupling laser beam is linearly polarized in the x-direction. For the PSEIT measurement, the $\lambda/4$ is used to make it circularly  polarized. The probe beam is further detected at DET1 and DET2.}
    \label{fig01}
\end{figure}

Figure \ref{fig01} (a) shows the configuration of the five-level atomic ladder used in this work ($5S_{1/2}$, $5P_{3/2}$, $68S_{1/2}$, $67P_{3/2}$ and $68P_{3/2}$) to measure the EIT-Autler Townes (AT) splitting to perform microwave electrometry. Here, we will compare different experimental techniques: i) the auxiliary microwave field \cite{auxiliary}; ii) the microwave amplitude modulation \cite{MWmodulation}; iii) and polarization spectroscopy \cite{DuarteGomes_2024,duarte2022polarization}. The basic common experimental setup, for all techniques, is composed of probe (red line) and coupling (blue line) laser beams, filtered by single-mode optical fibers, which travel in opposite directions and are focused at the center of an Rb cell (Fig. \ref{fig01} (b)). The probe laser beam operates at a calculated Rabi frequency $\Omega_{p}/2\pi = 2.3$ MHz and is stabilized to an optical cavity at the $^{85}$Rb  $5S_{1/2}(F=3) \rightarrow 5P_{3/2} (F=4)$ transition \cite{RodriguezFernandez2023}. The coupling laser beam has a calculated Rabi frequency of $\Omega_{c}/2\pi = 3.6$ MHz, and it is stabilized to the same optical cavity. Its frequency is scanned over the $5P_{3/2}(F=4) \rightarrow 68S_{1/2} (F=3)$ transition by varying the electro-optical modulator frequency (EOM) used in the optical cavity locking system \cite{RodriguezFernandez2023}. The probe beam is separated by a dichroic mirror (DM), and it passes through a half-wave liquid crystal retarder (LCR) (Thorlabs model LCC1111-B) and a polarizing beam splitter (PBS), which decomposes it into two orthogonal linearly polarized beams. Each beam is detected by a photodiode (DET1 and DET2). To detect the EIT signal for the auxiliary microwave field and the microwave amplitude modulation techniques, the probe and the coupling laser beams are linearly polarized in the x-direction, while the MW electric field is polarized in the z-direction, and a single detector detects the probe laser. To detect the polarization spectroscopy signal\cite{duarte2022polarization,DuarteGomes_2024}, both the probe and microwave fields are linearly polarized as before, while the coupling laser beam is circularly polarized. In this case, the photodiode signals are subtracted, resulting in a dispersive signal. The microwave (MW) is provided by a commercial two-channel MW generator (Model SynthPRO, Windfreak Technologies). Both channels are fed into an horn antenna (model PE9856B/SF-15 Pasternack) using a MW combiner (model ZX10-2-183-S+, Minicircuits). To ensure a plane wave on the Rb cell, the MW horn is placed 82 cm from it. One channel is tuned to the $68S_{1/2} \rightarrow 68P_{3/2}$ transition at 11.6660 $\pm$ 0.0001 GHz and is defined as the auxiliary MW field ($\Omega_{Aux}$). This MW field is used in the auxiliary microwave-dressed Rydberg EIT-AT technique \cite{auxiliary}. The other channel is tuned to the $68S_{1/2} \rightarrow 67P_{3/2}$ transition at 12.4550 $\pm$ 0.0001 GHz, and it is the target MW field ($\Omega_{MW}$). To increase the signal-to-noise ratio, the coupling beam is modulated at 1 kHz for the auxiliary microwave field \cite{auxiliary} and polarization spectroscopy techniques \cite{DuarteGomes_2024,duarte2022polarization}. For the MW amplitude modulation technique, an MW switch (model D1956, General Microwave) is used to modulate the target MW field at 1 kHz. The detected signals are processed by a lock-in amplifier. A zero-order vortex half wave retarder (VWR) (Thorlabs, model WPV10L-405) is used to generate the $LG_{0}^{1}$ mode with an efficiency of 97\%. Ten spectra were acquired for each experimental condition.

\section{Results}\label{Results}

\begin{figure}
    \centering
    \includegraphics[width=0.6\columnwidth]{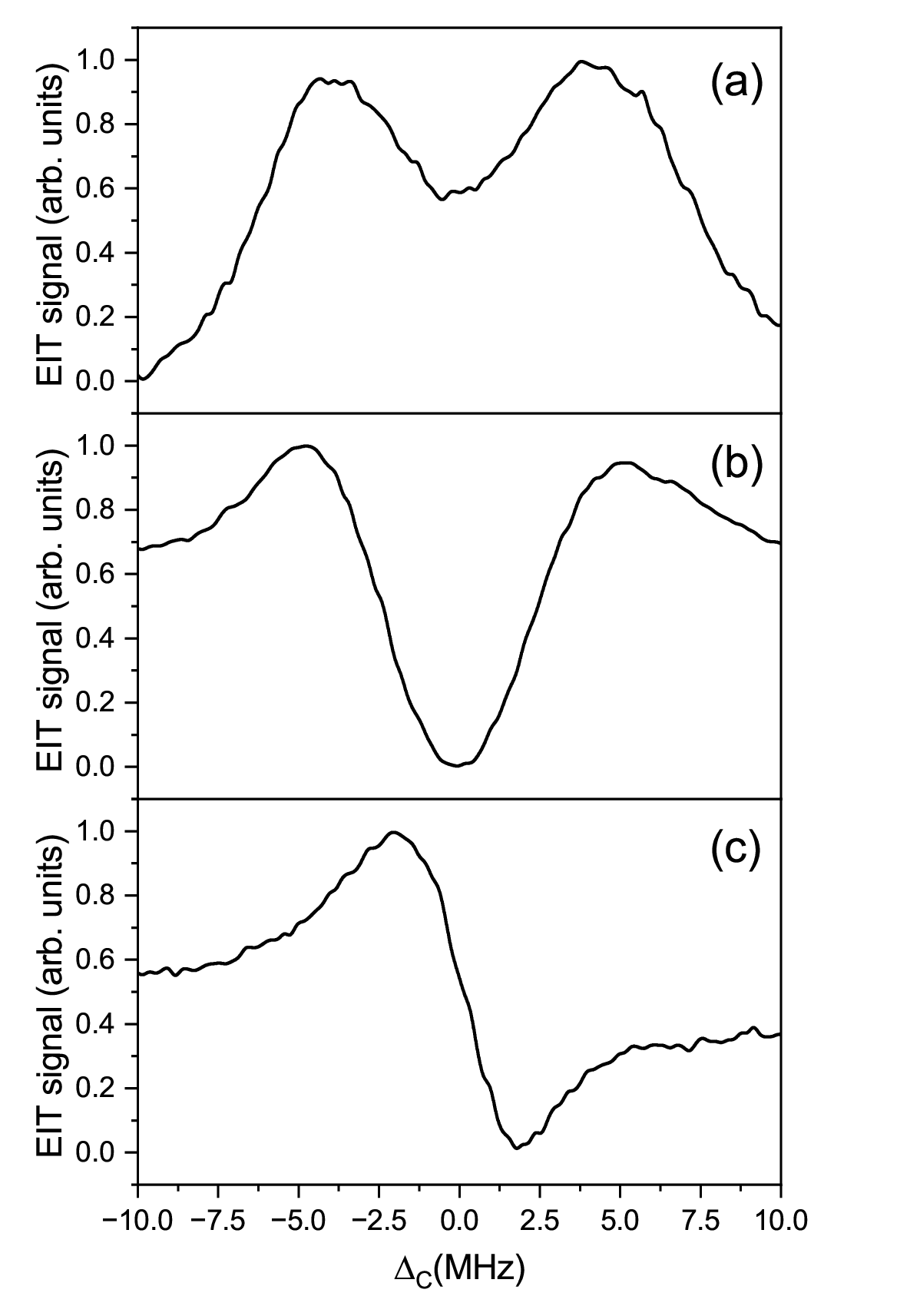}
    \caption{Typical EIT spectra as functions of coupling laser detuning ($\Delta_c$) for: (a) the auxiliary microwave field technique; (b) the microwave amplitude modulation technique; and (c) polarization spectroscopy technique. All curves were obtained at a 0.0001 mW MW power.}
    \label{fig02}
\end{figure}

Figure \ref{fig02} shows typical EIT spectra as functions of coupling laser detuning ($\Delta_c$) for the auxiliary microwave field technique (Fig. \ref{fig02}(a)); the microwave amplitude modulation technique (Fig. \ref{fig02}(b)); and polarization spectroscopy technique (Fig.\ref{fig02}(c)). For a fair comparison of both techniques, we processed all signals identically. For the data analysis, we employed the Python library SciPy \cite{2020SciPy-NMeth}. We used the \texttt{interp1d(x, y\_smooth, kind="cubic")} function to perform cubic spline interpolation. This method generates a set of cubic polynomials, each of which approximates the data between two consecutive points. The resulting interpolated curve is smooth and ensures continuity not only of the values but also of the first and second derivatives between the data points. By applying this technique, we obtain a smooth and continuous approximation of the experimental data, suitable for estimating values at points that lie between the observed measurements. Figure 3 shows an example for each technique, where the black curves correspond to the experimental data, and the red curves represent the interpolation polynomials. Following this, we identified the positions of the peaks by finding the minima of the interpolated polynomial using the function \texttt{minimize\_scalar(objective, bounds=bounds)} from the same library. This approach allowed us to accurately determine the critical points of the data. Here we have used the AT splitting defined in each different work: (a) $\Delta f_{m}$ from \cite{auxiliary}; (b) $\Delta f_{sho}$ from \cite{MWmodulation2} and (c) $\Delta_{PSEIT-AT}$ from \cite{DuarteGomes_2024}. We must point out that instead of using the zero crossing points of the EIT dispersion signal ($\Delta f_{AM}$), used by Liu et al. \cite{MWmodulation}, we used the two symmetric peaks proposed by Hao et al. \cite{MWmodulation2} ($\Delta f_{sho}$). According to the authors, this parameter allows the measurement of MW fields smaller than those of the zero-crossing parameter. The experimental points are the average of 10 spectra for each experimental condition, and the error bars are their standard deviation.

\begin{figure}
\centering
\includegraphics[width=0.6\columnwidth]{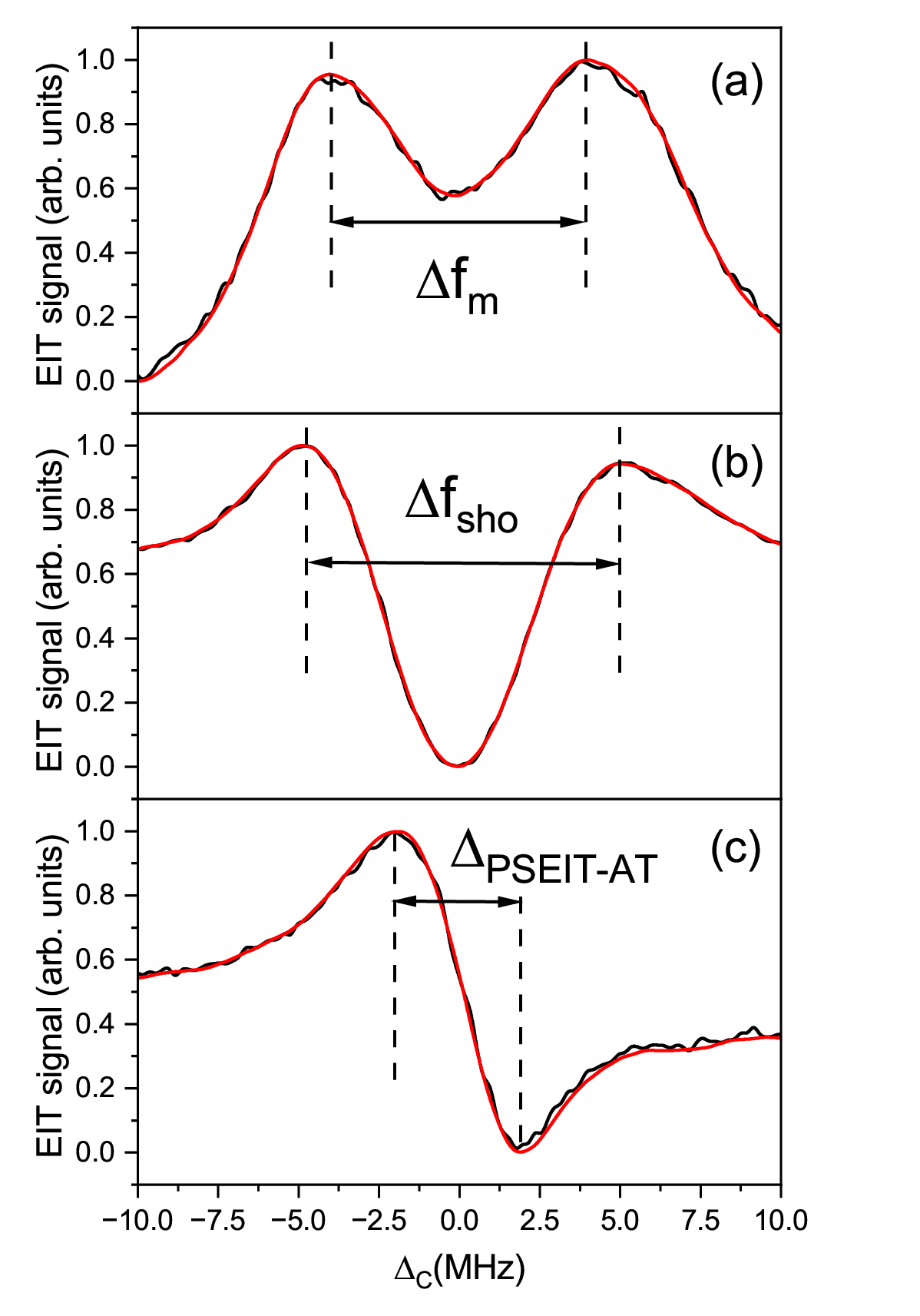}
\caption{(Color online) Typical EIT spectra as functions of coupling laser detuning ($\Delta_c$) for experimental curve (black lines) and interpolation curve (red lines). We have also pointed out the AT splitting used in each technique: (a) $\Delta f_m$ for the auxiliary microwave field technique \cite{auxiliary}; (b) $\Delta f_{sho}$ the microwave amplitude modulation technique \cite{MWmodulation2}; and (c) $\Delta_{PSEIT-AT}$ for the polarization spectroscopy technique \cite{DuarteGomes_2024}.}\label{fig03}
\end{figure}

\begin{figure}
\centering
\includegraphics[width=0.6\columnwidth]{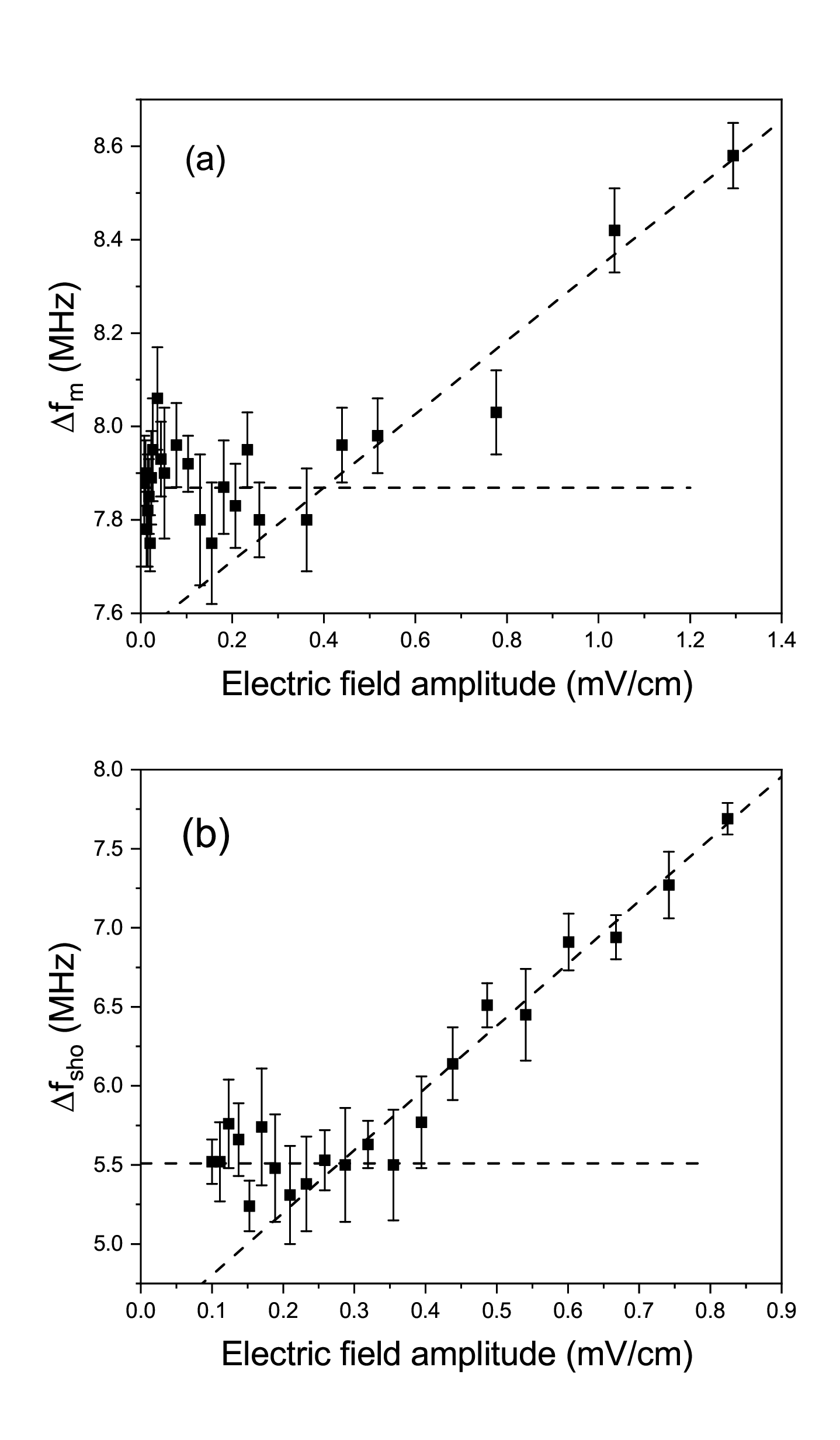}
\caption{ (a) $\Delta f_m$ and (b) $\Delta f_{sho}$ as a function of the amplitude of the electric field. The lines are two linear fit functions, whose intersection, allows the determination of the minimum measurable MW electric field amplitude, which is $0.40 \pm 0.02$ mV/cm and $0.27 \pm 0.02$ mV/cm for the auxiliary microwave field technique and the microwave amplitude modulation technique, respectively.The experimental points are an average of 10 spectra, and the error bars are their standard deviation.  }\label{fig04}
\end{figure}

\begin{figure}
\centering
\includegraphics[width=0.6\columnwidth]{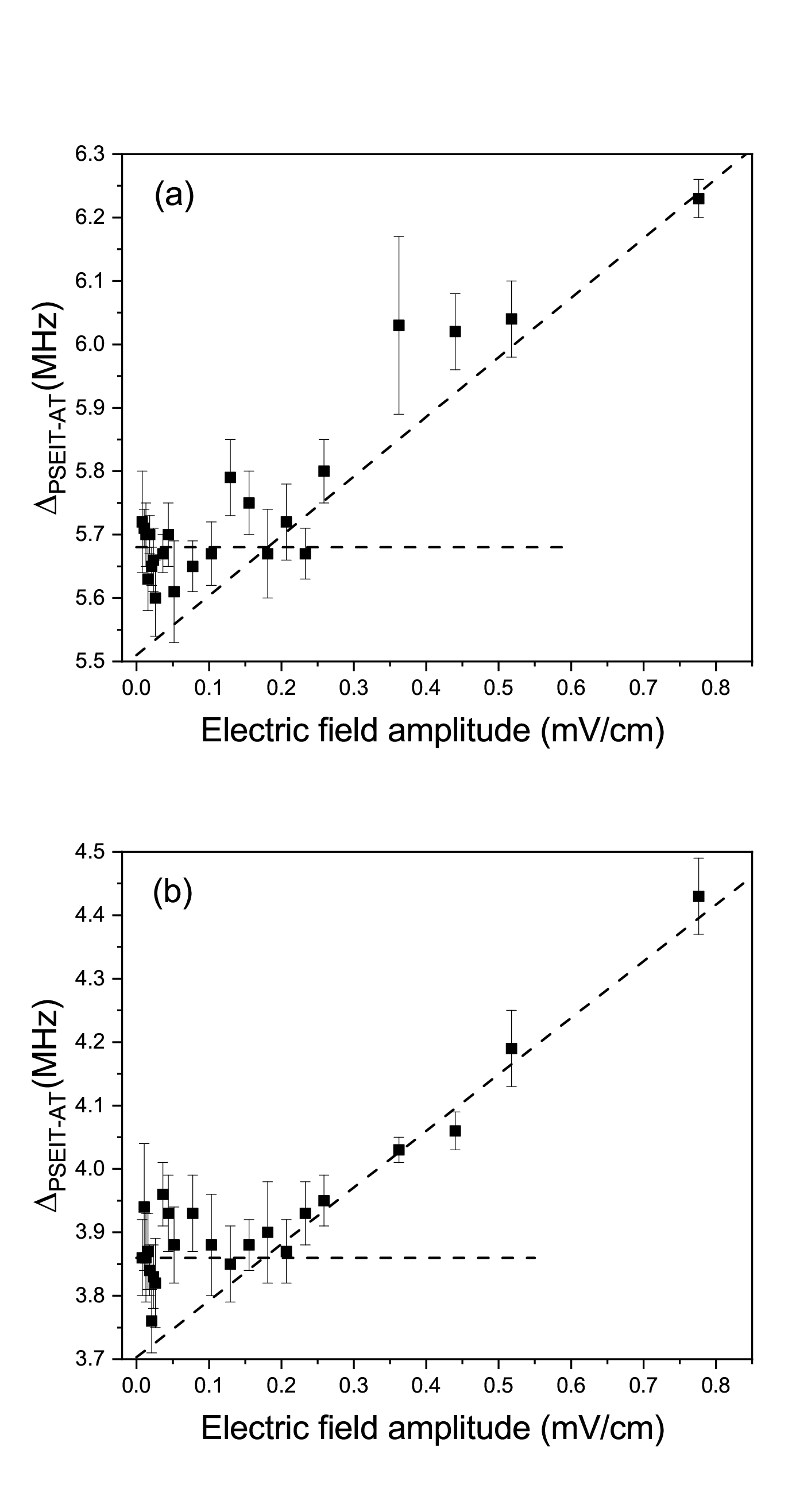}
\caption{$\Delta_{PSEIT-AT}$ as functions of the applied electric field using polarization spectroscopy with (a) Gaussian and (b) Laguerre-Gauss coupling laser beam. The minimum measurable MW electric field amplitude is $0.18 \pm 0.02$ mV/cm for Gaussian coupling beam and $0.17 \pm 0.02$ mV/cm  for the Laguerre-Gauss coupling beam. The experimental points are an average of 10 spectra, and the error bars are their standard deviation.  }\label{fig05}
\end{figure}

To calibrate the MW field, we have measured the AT splitting in the linear regime as a function of the root square of the MW power. 
This was done for the auxiliary MW field ($\Omega_{Aux}$ at the
$68S_{1/2} \rightarrow 68P_{3/2}$ transition at 11.666 GHz) and the target MW field ($\Omega_{MW}$ at the $68S_{1/2} \rightarrow 67P_{3/2}$ transition at 12.455 GHz). Figure \ref{fig04}(a) shows $\Delta f_m$ as a function of the amplitude of the electric field, which allows measurement of the amplitude of the minimum MW electric field. For a low field, $\Delta f_m$ is constant; as the field increases, $\Delta f_m$ also increases. The intersection between two linear fit functions allows the determination of the minimum measurable MW electric field amplitude, which is $0.40 \pm 0.02$ mV/cm for the auxiliary microwave field technique. Figure \ref{fig04}(b) shows $\Delta f_{sho}$ as a function of the amplitude of the electric field, which allows us to determine the minimum measurable MW electric field amplitude of $0.27 \pm 0.02$ mV/cm for the microwave amplitude modulation technique. Figures \ref{fig05} (a) and (b) show $\Delta_{PSEIT-AT}$ as functions of the applied electric field using polarization spectroscopy with the Gaussian and Laguerre-Gauss coupling laser beam, respectively. For the Gaussian coupling beam, we have measured a minimum measurable MW electric field amplitude of $0.18 \pm 0.02$ mV/cm. For the Laguerre-Gauss coupling beam, we have measured $0.17 \pm 0.02$ mV/cm.

\section{Discussion}\label{Discussion}

In our work, the polarization spectroscopy technique obtained the best results, with a minimum measurable MW electric field amplitude of about $0.18 \pm 0.02$ mV/cm. The difference between the Gaussian and Laguerre-Gaussian coupling laser beams was negligible. The only difference is that the $\Delta_{PSEIT-AT}$ value for zero-MW power was smaller for the Laguerre-Gaussian coupling laser beam than for the Gaussian coupling laser beam. We must point out that our results are $\simeq 4$ times better than those obtained by Gomes et al. \cite{DuarteGomes_2024}. The two linear fit functions, in a linear plot, allow for a smaller minimum measurable MW electric field amplitude than in a log-log plot. We believe this is due to the fact we have used a smaller electric field fitting range in this work ($\leq 1 mV/cm$).
Our result for the minimum measurable MW electric field amplitude using the auxiliary field technique ($0.40 \pm 0.02$ mV/cm) is worse than the result obtained by Jia et al. (31 $\mu$ V/cm) \cite{auxiliary}. We should point out that a direct comparison is difficult for several reasons: i) The authors have measured only six points in a small MW field (< 1 mV/cm); ii) The origin of the error bars is not discussed in their work; iii) They did not describe the procedure used to measure $\Delta f_{m}$.  

Our result for the minimum measurable MW electric field amplitude using the microwave amplitude modulation technique ($0.27 \pm 0.02$ mV/cm) is better than the result obtained by Liu et al. (430 $\mu$V/cm) \cite{MWmodulation}. However, it is important to note that we have used $\Delta f_{sho}$ \cite{MWmodulation2} instead of $\Delta f_{AM}$ \cite{MWmodulation}. On the other hand, our result is worse than that obtained by Hao et al. (56 $\mu$V/cm at 9.2 GHz). Again, direct comparison is difficult for several reasons: i) the authors have measured only a few points in a small MW field (< 1 mV/cm); ii) The origin of the error bars is not discussed in their work; iii) They did not describe the procedure used to measure $\Delta f_{sho}$.

\section{Conclusions}\label{Conclusion}

In this work, we have compared different Rydberg atom-based microwave electrometry techniques (auxiliary microwave field \cite{auxiliary}; microwave amplitude modulation \cite{MWmodulation}; and polarization spectroscopy \cite{DuarteGomes_2024, duarte2022polarization}) for the same atomic Rydberg states and under the same experimental conditions. Our results indicate that polarization spectroscopy allows for the best minimum measurable MW electric field amplitude ($0.18 \pm 0.02$ mV/cm). It is essential to highlight that all techniques, used in this work, require calibration of the AT splitting against a known microwave electric field prior to its usage for measuring an unknown field. Consequently, it may serve as a secondary standard.  We should also point out that the use of a microwave lens could expand its range of applications, as demonstrated in \cite{DuarteGomes_2024}.

\section*{Funding}
This work is supported by grants 2019/10971-0 and 2021/06371-7, S\~{a}o Paulo Research Foundation (FAPESP), and CNPq (305257/2022-6). It was supported by the Army Research Office - Grant Number W911NF-21-1-0211.

\clearpage

\bibliographystyle{unsrt} 
\bibliography{References} 

\end{document}